\documentclass[prd,aps,preprint,tightenlines,showpacs,superscriptaddress,nofootinbib]{revtex4-1}

\usepackage{bbm}
\usepackage{mathrsfs}
\usepackage{amsfonts}
\usepackage{amsmath}
\usepackage{array}
\usepackage{verbatim}
\usepackage{epsfig}
\usepackage{graphicx}
\newcommand{\bqa}{\begin{eqnarray}}
\newcommand{\eqa}{\end{eqnarray}}

\def\slash{\llap /}
\def    \q              {\ifmmode {\cal{Q}} \else ${\cal{Q}}$ \fi}
\def    \nn             {\nonumber}
\def    \=              {\;=\;}
\def\slash#1{{#1\!\!\!/}}
\def    \o              {\ifmmode {\cal{O}} \else ${\cal{O}}$ \fi}

\begin{document}

\title{Universal Suppression of Heavy Quarkonium Production in $pA$ Collisions
at Low Transverse Momentum}

\author{Jian-Wei Qiu}
\affiliation{Physics Department, Building 510A, Brookhaven National Laboratory, Upton, NY 11973, USA}

\author{Peng Sun}
\affiliation{Nuclear Science Division, Lawrence Berkeley National
Laboratory, Berkeley, CA 94720, USA}

\author{Bo-Wen Xiao}
\affiliation{Institute of Particle Physics, Central China Normal University, Wuhan 430079, China}

\author{Feng Yuan}
\affiliation{Nuclear Science Division, Lawrence Berkeley National
Laboratory, Berkeley, CA 94720, USA}

\begin{abstract}
The nuclear suppression of heavy quarkonium production at low
transverse momentum in $pA$ collisions in high energy scatterings
is investigated in the small-$x$ factorization formalism. 
A universal suppression is found in the large $N_c$ limit between
the two formalisms to describe the heavy quarkonium production:
the non-relativistic QCD (NRQCD) and the color-evaporation model (CEM). 
This provides an important probe to the saturation momentum at small-$x$
in big nucleus. We also comment on the phenomenological applications
of our results.
\end{abstract}
\maketitle

\section{Introduction}

Gluon saturation at small-$x$ in a large nucleus has attracted great attention 
in the last few decades, and progresses have been made 
theoretically~\cite{Gribov:1984tu,Mueller:1985wy,McLerran:1993ni,arXiv:1002.0333, eic}.
The emergence of this phenomenon  changes
the landscape of parton distributions inside the nucleon/nucleus.
In the dense region, the characteristics of QCD dynamics is different
from that in the dilute region, for example, the
evolution equation has to be modified to account
for coherent interactions among high density gluons. 
To better explore the strong interaction dynamics in the dense region, 
it is important to have physical processes with two distinct scales, 
one hard scale $Q$ to localize the probe and a relatively soft scale 
to be sensitive to the saturation physics~\cite{Accardi:2012qut}.
It was realized that two-scale observables make it possible to perform
the small-$x$ calculations in terms of the factorization approach~\cite{Dominguez:2010xd}, 
and to provide directly access to the unintegrated gluon distributions (UGDs), 
which are important ingredients in the saturation physics, and 
unveil the importance of the coherent multiple interaction effects 
in the small-$x$ calculations~\cite{arXiv:1002.0333}.
Several processes have been proposed in the literature, including
semi-inclusive DIS~\cite{Marquet:2009ca}, low $P_\perp$
Drell-Yan~\cite{Baier:2004tj}, and back-to-back di-hadron
correlations in forward $pA$ processes~\cite{Dominguez:2010xd,JalilianMarian:2004da,Blaizot:2004wv,Marquet:2007vb,Albacete:2010pg,Stasto:2011ru,Deak:2011ga,Kutak:2012rf,Lappi:2012nh,Dominguez:2012ad,Iancu:2013dta}.

In this paper, we will demonstrate that the heavy quarkonium production in
the forward region of $pA$ collisions is an excellent probe of the saturation physics 
at small-$x$.  Earlier works to calculate the heavy quarkonium production using
the saturation formalism have been done in Refs.~\cite{Kharzeev:2005zr,Kharzeev:2008cv,Kharzeev:2008nw,Dominguez:2011cy,Kharzeev:2012py}, where the inclusive productions
are the main focus and the calculations are done in the simple color-singlet model.  
In our calculations below, we concentrate on rapidity and transverse momentum distribution, 
and follow an effective factorization approach~\cite{Dominguez:2010xd}
with two models dealing with the hadronization from heavy quark pairs 
to physical quarkonia.
One model is based on non-relativistic QCD (NRQCD)~\cite{Bodwin:1994jh}, 
an effective theory of QCD, and the other
is the so-called color evaporation model (CEM)~\cite{Fritzsch:1977ay}.
Both NRQCD model and CEM share a similar idea for heavy quarkonium production:\ 
heavy quark pairs are produced at short distance, and all pairs, at various spin states
with/without color, could hadronize into physical heavy quarkonia. 
In NRQCD, the hadronization probability from different states of heavy
quark pairs is organized by the power of $\alpha_s$ and the power of 
heavy quark velocity $v$ in the pair's rest frame.  In CEM, on the other hand, 
every produced heavy quark pairs have the same probability to transform
into the same quarkonium state, so long as the pairs' invariant masses 
are within the open charm threshold.  Both approaches give a reasonable
description of unpolarized $J/\psi$ and $\Upsilon$ production at 
a large transverse momentum at collider energies
\cite{Brambilla:2004wf}.

\begin{figure}[tbp]
\centering

\includegraphics[width=8cm]{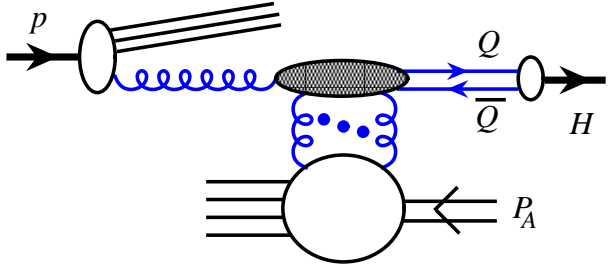}

\caption{Sketch for heavy quarkonium production at a forward rapidity in $pA$ collisions.}
\label{fig:pA2QQb}
\end{figure}

We focus our calculations of heavy quarkonium production in $pA$ collisions 
in the forward region (large rapidity region along the direction of proton), 
as sketched in Fig.~\ref{fig:pA2QQb}, 
where we believe that a factorization approach might be justified.  In addition, 
we concentrate on the production of heavy quarkonium with relatively
low transverse momentum $P_\perp$, which is mainly generated by 
coherent multiple scattering of soft gluons from the nuclei.  
More specifically, we present our calculations for the region,  
$\Lambda_{\rm QCD} \ll P_\perp \sim Q_s(A) \ll M$, where 
$M$ is quarkonium mass and $Q_s(A)$ is 
the saturation scale of the nuclei with atomic weight $A$.  
In general, the factorization between the production of heavy quark pair
and the pair's hadronization to a physical quarkonium, 
assumed to be true in both NRQCD and CEM, 
is not necessarily valid in this kinematic regime.
When $Q_s(A) \sim m v$ with heavy quark mass $m \sim M/2$, 
multiple scattering of heavy quark pair with gluons from nuclei
could interfere with the pair's hadronization process, and invalidate 
the velocity expansion of the NRQCD approach.  However,
in the very forward region, the system of heavy quark pair moves through 
nuclear matter with a very large longitudinal momentum $P_\parallel$, 
the dynamics of coherent interaction between the pair, which leads to the 
formation of a bound quarkonium, is strongly time dilated.  
With a time dilation factor $P_\parallel/M$, we conclude that if
\begin{equation}
\frac{1}{mv} \left(\frac{P_\parallel}{M}\right) 
\gg \frac{1}{P_\perp}  \sim \frac{1}{Q_s(A)}
\ \ \Leftrightarrow  \ \
y \gg \ln\left(\frac{2mv}{P_T}\right) 
\sim \ln\left(\frac{Mv}{Q_s(A)}\right) ,
\label{eq:fac-condition}
\end{equation}
the pair's hadronization is effectively frozen (or not even started) 
when the pair passes through and interacts with the nuclei,
so that the pair's hadronization is effectively factorized from the 
production of the pair.  

The condition, $\Lambda_{\rm QCD} \ll P_\perp$, effectively ensures 
that the interaction between the heavy quark pair and the 
spectators of incoming proton is suppressed by powers of $Q_s(p)/P_\perp$
with proton saturation scale $Q_s(p)\sim \Lambda_{\rm QCD}$ in this region, and 
we could use collinear factorization to factorize the active gluon
from the colliding proton.  The condition in Eq.~(\ref{eq:fac-condition}) allows us to 
calculate the multiple interactions of the pair with the nuclear target. 
It was shown~\cite{Dominguez:2010xd} that for hard processes in $pA$ collisions, 
these multiple interactions could be reduced to the un-integrated gluon distributions 
which can be easily formulated in small-$x$ saturation formalism.
We show that such an effective $k_\perp$ factorization from the nucleus 
is also valid for heavy quarkonium production in $pA$ collisions, 
where a very important feature emergences that the un-integrated gluon distribution
is the same in both NRQCD and CEM approaches in the
large $N_c$ limit. Therefore, these two approaches will give the same
predictions for the nuclear suppression factor in our calculations.

When $P_\perp \ll M$, the transverse momentum distribution of the pair
in strongly influenced by the initial-state gluon shower, which leads to 
$\alpha_s^m\ln^{2m-1}(M^2/P_\perp^2)$ type Sudakov large logarithms 
in perturbative calculation of the production rate~\cite{Sudakov:1954sw}.  
Such large logarithmic contributions have to be resummed.   
We use the Collins-Soper-Sterman (CSS) formalism 
for the resummation~\cite{Collins:1984kg}. 
For heavy quarkonium production in hadronic collisions,
resummation of Sudakov large logarithms was first addressed in
Ref.~\cite{Berger:2004cc} by using CSS formalism and CEM for the formation 
of quarkonia.  Recently, the same resummation were carried out in the 
NRQCD formalism as well~\cite{Sun:2012vc}.

It was demonstrated recently that the soft gluon resummation and small-$x$ evolution
can be performed consistently for the hard processes in $pA$ collisions~\cite{Mueller:2012uf}. 
We apply the same technique to heavy quarkonium production in $pA$ collisions. 
The combination of the QCD resummation and the effective $k_\perp$-factorization 
from the nucleus in $pA$ collisions at small-$x$ makes the heavy quarkonium production 
as a powerfull tool to investigate the saturation physics. 
Most importantly, we show that the hadronization mechanism 
does not affect the sensitivity of the nuclear suppression of 
heavy quarkonium production in the forward $pA$ collisions.

The rest of the paper is organized as the following. In
Sec.~II, we formulate the heavy quarkonium production
at low transverse momentum in the small-$x$ formalism
and NRQCD.  An effective $k_\perp$ factorization emerges
for heavy quarkonium production with two different 
un-integrated gluon distributions in the NRQCD formalism. 
We extend the Sudakov
resummation to heavy quarkonium production in Sec.~III,
and discuss heavy quarkonium suppression in $pA$ collisions in Sec.~IV.  
Finally, we summarize our paper and comment on the 
phenomenological applications in Sec.~V.

\section{Unintegrated Gluon Distributions in Heavy Quarkonium
Production in $pA$ Collisions}

In this section, we calculate the heavy quarkonium production
at low transverse momentum
in $pA$ collisions in the small-$x$ factorization formalism, and
demonstrate that an effective $k_\perp$-factorization applies for
both NRQCD and CEM models. We take
$P_\perp\ll M_Q$ limit, and neglect all higher order corrections.
Since the details in production mechanism is different in NRQCD and
CEM, we present the calculations separately.

Let us recall the two basic unintegrated gluon distributions in the
small-$x$ formalism. The first gluon distribution, the so-called
Weizs\"acker-Williams (WW) gluon distribution can be defined following
the conventional gluon distribution~\cite{Collins:1981uw,{Ji:2005nu}}
\begin{eqnarray}
xG^{(1)}(x,k_{\perp })&=&\int \frac{d\xi ^{-}d^2\xi _{\perp }}{(2\pi
)^{3}P^{+}}e^{ixP^{+}\xi ^{-}-ik_{\perp }\cdot \xi _{\perp }}   \langle P|F^{+i}(\xi ^{-},\xi _{\perp })\mathcal{L}_{\xi }^{\dagger
}\mathcal{L}_{0}F^{+i}(0)|P\rangle \ ,  \label{g1}
\end{eqnarray}%
where $F^{\mu \nu }$ is the gauge field strength tensor $F_a^{\mu \nu
}=\partial ^{\mu }A_a^{\nu }-\partial ^{\nu }A_a^{\mu }-gf_{abc}A_b^\mu
A_c^\nu$ with $f_{abc}$ the antisymmetric structure constants for $SU(3)$ 
with $a,b,c,$ the color indices, and $\mathcal{L}_{\xi }$
is the gauge link in the adjoint representation $A^{\mu}=A_{a}^{\mu }t_{a}$ with $t_{a}=-if_{abc}$.
In the small-$x$ factorization calculations, this distribution can be written in terms of
the correlator of four Wilson lines as~\cite{Dominguez:2010xd},
\begin{equation}
xG^{(1)}(x,k_\perp)=-\frac{2}{\alpha_S}\int\frac{d^2v}{(2\pi)^2}\frac{d^2v'}{(2\pi)^2}\;e^{-ik_\perp\cdot(v-v')}\left\langle\text{Tr}\left[\partial_iU(v)\right]U^\dagger(v')\left[\partial_iU(v')\right]U^\dagger(v)\right\rangle_{x_g},
\end{equation}
where the Wilson line $U(x_{\perp})$ is defined as
$U^n\left[-\infty,+\infty; x_{\perp}\right]$, and the notation
$\langle \dots \rangle_{x_g}$ is used for the Color Glass Condensate (CGC) 
average of the color charges over the nuclear wave function 
where $x_g$ is the smallest fraction of longitudinal momentum probed, and is
determined by the kinematics.
For a large nucleus, this distribution at a small-$x$ can be evaluated using the
McLerran-Venugopalan model~\cite{McLerran:1993ni}
\begin{equation}
xG^{(1)}(x,k_{\perp })=\frac{S_{\perp }}{\pi ^{2}\alpha _{s}}\frac{%
N_{c}^{2}-1}{N_{c}}\int \frac{d^{2}r_{\perp }}{(2\pi )^{2}}\frac{%
e^{-ik_{\perp }\cdot r_{\perp }}}{r_{\perp }^{2}}\left( 1-e^{-\frac{r_{\perp
}^{2}Q_{s}^{2}}{4}}\right) \ ,
\label{mvmodel}
\end{equation}%
where $N_c=3$ is the number of colors, $S_\perp$ is the transverse area of the target nucleus, and $Q^2_{s}=\frac{g^2N_c}{4\pi}\ln\frac{1}{r_{\perp}^2\lambda^2}\int dx^{-} \mu^2(x^{-})$ is the gluon saturation
scale with $\mu^2$ the color charge density in a large nuclei.

The second gluon distribution, the Fourier transform of the dipole
cross section, is defined in the fundamental
representation,
\begin{eqnarray}
xG^{(2)}(x,k_{\perp }) &=&2\int \frac{d\xi ^{-}d\xi _{\perp }}{(2\pi
)^{3}P^{+}}e^{ixP^{+}\xi ^{-}-ik_{\perp }\cdot \xi _{\perp }}\langle P|\text{%
Tr}\left[F^{+i}(\xi ^{-},\xi _{\perp })\mathcal{U}^{[-]\dagger }F^{+i}(0)%
\mathcal{U}^{[+]}\right]|P\rangle \ ,  \label{g2}
\end{eqnarray}
where the gauge link
$\mathcal{U}_\xi^{[-]}=U^n\left[0,-\infty;0\right]U^n\left[-\infty,
\xi^{-}; \xi_{\perp}\right]$ stands for initial state
interactions. Thus, the dipole gluon distribution contains both
initial and final state interactions in the definition, and
can be calculated in the MV model,
\begin{equation}
xG^{(2)}(x,q_\perp)= \frac{q_{\perp }^{2}N_{c}}{2\pi^2 \alpha_s}%
S_{\perp }\int
\frac{d^2r_\perp}{(2\pi)^2}e^{-iq_\perp\cdot r_\perp}
\frac{1}{N_c}\left\langle\text{Tr}U(0)U^\dagger(r_\perp)\right\rangle_{x_g}\equiv  \frac{q_{\perp }^{2}N_{c}}{2\pi^2 \alpha_s}%
S_{\perp } F(q_\perp).
\end{equation}

\subsection{NRQCD Calculations}

In the NRQCD formalism, we can write down the production of heavy quarkonium
into the following form,
\begin{eqnarray}
d\sigma=\sum_n d\hat{\sigma}(Q\bar{Q}[n]+X)\langle\o^H[n]\rangle \; .
\end{eqnarray}
Here $\hat{\sigma}(Q\bar{Q}[n]+X)$ is the cross section at the
parton level, which represents the production of a pair of heavy
quark in a fixed color, spin and orbital angular momentum state
$n$ and can be calculated perturbatively. The long distant matrix
element (LDME) $\langle\o^H[n]\rangle$ describes the transition of
the heavy quark pair in the configuration of $Q\bar{Q}[n]$ into
the final state heavy quarkonium.  LDMEs are organized in terms of the velocity $v$
expansion in the NRQCD framework. For $J/\psi$ production in $pp$
collisions, the differential cross sections depend on the
following three color-octet matrix elements:
\begin{equation}
\langle\mathcal {O}^{J/\psi}[{}^3S_1^8)]\rangle\; ,\langle\mathcal{O}^{J/\psi}[{}^1S_0^8]\rangle\; ,\langle\mathcal
{O}^{J/\psi}[{}^3P_J^8]\rangle \ ,
\end{equation}
which are at the same order in the velocity expansion.

\begin{figure}[tbp]
\centering

\includegraphics[width=9cm]{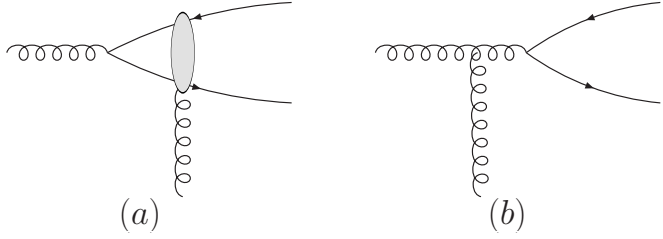}

\caption{Multiple interaction in heavy quarkonium production in $pA$ collisions:
(a) multiple interactions between the heavy quark pair and the nucleus target;
(b) multiple interactions between the incoming gluon and the nucleus target.}
\label{fig:rhic}
\end{figure}

In the following, we will take the example for ${}^1S_0$ channel in
both color-singlet and color-octet configurations to show
how the unintegrated gluon distributions enter into the
factorization formalism at low transverse momentum. The typical diagrams
are plotted in Fig.~2, where the final state $Q\bar Q$ can be
either color-singlet or color-octet.
According to the NRQCD factorization, we calculate the
perturbative part of heavy quark pair production, and project
out different spin and color configuration.
Using the spin projection we listed in the Appendix,
we find that the amplitude for ${}^1S_0$ can be written as
\begin{eqnarray}
\mathcal {M}^\mu( {}^1S_0^8)
&=&-\int \frac{d^2k_{g1\perp}}{(2\pi)^2}d^2x_{1\perp}d^2x_{2\perp}e^{ik_{g1\perp}\cdot x_{1\perp}+ik_{g2\perp}\cdot x_{2\perp}}\nonumber\\
&&\frac{16  m_c \epsilon ^{\mu (k_{g1}-k_{g2})np
}}{4m_c^2-(k_{g1\perp}-k_{g2\perp})^2}\frac{1}{\sqrt{4m_c^3}}
2\pi\delta(k_g^-)(\textrm{Tr}[T^bU_{k_{g1}}(x_{1\perp})T^aU_{k_{g2}}^\dag(x_{2\perp})]) \ ,\\
\mathcal {M}^\mu( {}^1S_0^1)
&=&-\int \frac{d^2k_{g1\perp}}{(2\pi)^2} d^2x_{1\perp}d^2x_{2\perp}e^{ik_{g1\perp}\cdot x_{1\perp}+ik_{g2\perp}\cdot x_{2\perp}}\nonumber\\
&&\frac{16  m_c \epsilon ^{\mu (k_{g1}-k_{g2})np
}}{4m_c^2-(k_{g1\perp}-k_{g2\perp})^2}\frac{1}{\sqrt{4m_c^3}}
2\pi\delta(k_g^-)(\textrm{Tr}[U_{k_{g1}}(x_{1\perp})T^aU_{k_{g2}}^\dag(x_{2\perp})]) \ ,
\end{eqnarray}
where $a$ is color index for incoming gluon, and $b$ for that
in the final state if it is a color-octet channel. In the above equations,
$U$ represent the multiple interaction with the nucleus target $U(x_\perp)=\mathcal{P}\exp\left\{ig_S\int_{-\infty}^{+\infty} \text{d}%
x^+\,T^cA_c^-(x^+,x_\perp)\right\}$.

First, let us examine the color-singlet channel. In the small transverse momentum
limit, $P_\perp\ll M_Q$, we have $k_{gi\perp}\sim P_{\perp}\ll M_Q$, and we can
expand the amplitude in terms of $k_{gi\perp}/M_Q$. In the leading order
expansion, there is no dependence on $k_{gi\perp}$ except the Fourier transformation
factor, which leads to $\delta ^2(x_{1\perp}-x_{2\perp})$. The latter equation
results into vanishing contribution, because ${\rm Tr}(T^a)=0$. Therefore, nonzero
leading order expansion is proportional to $k_{g1\perp}^\alpha-k_{g2\perp}^\alpha$,
and its contribution leads to $\partial ^\alpha_\perp U^\dagger(x_\perp) U(x_\perp)$.
This is exactly the WW gluon distribution. It is consistent to the color-single, because
there is only the initial state interaction effects, and the gluon distribution is the same
as that in the Higgs boson production process.

For the the color-octet channel, however, it has final state interaction as well. The UGD
will be different from WW gluon distribution. Again, if we work at the leading nonzero
order, we find that, in the large $N_c$ limit, the relevant gluon distribution ${\cal F}^{(c\bar c)[8]}$ is 
\begin{equation}
\frac{S_{\perp }}{\pi ^{2}\alpha _{s}}\frac{
N_{c}^{2}-1}{4N_{c}}\int d^2q_1 d^2 q_2 \delta^{(2)} (P_\perp-q_1-q_2) F(q_1)F(q_2) (q_1-q_2)^2 \ .
\end{equation}
Or in terms of the notation of Ref.~\cite{Dominguez:2010xd},
\begin{eqnarray}
{\cal F}^{(c\bar c)[8]}={\cal F}^{(1)}_{gg}+{\cal F}^{(2)}_{gg} \ ,
\label{eq:WWgluon}
\end{eqnarray}
where ${\cal F}^{(1,2)}_{gg}$ are defined in Ref.~\cite{Dominguez:2010xd}.

Now, we turn to ${}^3S_1$ channel. It is different from ${}^1S_0$ channel,
there will be additional diagram in Fig.~2(b) contributing to color-octet
${}^3S_1$ channel. In the following, we will show, however, that the ${}^3S_1$
channel is power suppressed both for the color-singlet and color-octet
contributions. This is consistent with the power counting in the collinear
factorization calculation in the NRQCD of Ref.~\cite{Sun:2012vc}.

First, let us write down the contribution from Fig.~2(a), which
reads,
\begin{eqnarray}
\mathcal {M}^{\rho\mu}( {}^3S_1^1)&=&\int \frac{d^2k_{g1\perp}}{(2\pi)^2}d^2x_{1\perp}d^2x_{2\perp}e^{ik_{g1\perp}\cdot x_{1\perp}+ik_{g2\perp}\cdot x_{2\perp}}\frac{1}{\sqrt{4m_c^3}}2\pi\delta(k_g^-)\nonumber\\
&&\frac{16 \mathbbm{i}m_c^2 \left(k_{g1}^{\mu } n^{\rho
}+k_{g2}^{\mu } n^{\rho }-2 g^{\mu \rho }
 n\cdot p\right)}{(k_{g1\perp}-k_{g2\perp})^2-4m_c^2}(\textrm{Tr}[U(x_{1\perp})T^aU^\dag(x_{2\perp})])\ ,
\end{eqnarray}
for the color-singlet channel. In the low transverse momentum region, the power counting analysis tells us that the above
vanishes in the leading power of $P_\perp/M_Q$. The first term is cancels out that
of the second term because of
$\partial_\perp^\alpha U^\dagger(x_\perp)U(x_\perp)=-U^\dagger\partial_\perp^\alpha U(x_\perp)$.
The third term is power suppressed because of ${\rm Tr}(T^a)=0$.

The color-octet ${}^3S_1$ can be analyzed in a similar manner. The contribution
from Fig.~2(a) can be written as,
\begin{eqnarray}
\mathcal {M}^{\rho\mu}( {}^3S_1^8)_{\rm fig.2(a)}&=&\int \frac{d^2k_{g1\perp}}{(2\pi)^2}d^2x_{1\perp}d^2x_{2\perp}e^{ik_{g1\perp}\cdot x_{1\perp}+ik_{g2\perp}\cdot x_{2\perp}}\frac{1}{\sqrt{4m_c^3}}2\pi\delta(k_g^-)\nonumber\\
&&\frac{16 \mathbbm{i}m_c^2 \left(k_{g1}^{\mu } n^{\rho
}+k_{g2}^{\mu } n^{\rho }-2 g^{\mu \rho }
 n\cdot p\right)}{(k_{g1\perp}-k_{g2\perp})^2-4m_c^2}(\textrm{Tr}[T^bU(x_{1\perp})T^aU^\dag(x_{2\perp})-T^bT^a])\ ,
\end{eqnarray}
while the contribution from Fig.~2(b) is
\begin{eqnarray}
\mathcal {M}^{\rho\mu}(
{}^3S_1^8)_{\rm fig.2(b)}&=&\int d^2x_{\perp}e^{iP_{\perp}\cdot x_{\perp}}\frac{1}{\sqrt{4m_c^3}}2\pi\delta(k_g^-)\nonumber\\
&&\frac{16 \mathbbm{i} m_c^2 \left(k_g^{\mu } n^{\rho }-2 g^{\mu
\rho } n\cdot p\right)}{-4
m_c^2}\frac{1}{\sqrt{4m_c^3}}(\textrm{Tr}[T^bU(x_{\perp})T^aU^\dag(x_{\perp})-T^bT^a])\ .
\end{eqnarray}
Clearly, the leading order expansions cancel out each other between these two
diagrams. Therefore, in the leading power, there is no contribution from ${}^3S_1$
channel in either color-singlet or color-octet channel. This again, is consistent
with the collinear factorization calculation in NRQCD.

We have also calculated the contributions from ${}^3P_J$ states in both color-singlet
and color-octet channels (see the detailed derivations in the Appendix), 
and we find that the non-vanishing contribution in the
color-singlet channel is proportional to the WW gluon distribution, whereas that
from the color-octet channel proportional to the same UGD as that for ${}^1S_0^8$
channel. Therefore, in summary, we can write down an effective $k_\perp$-factorization
for heavy quarkonium production in low $P_\perp$ region as
\begin{eqnarray}
\frac{d\sigma^{(1)}}{dyd^2P_\perp}&=&\sigma_0^{(1)}xg_p(x) x_AG^{(1)}(x_A,P_\perp) \ , \\
\frac{d\sigma^{(8)}}{dyd^2P_\perp}&=&\sigma_0^{(8)}xg_p(x) {\cal F}^{(c\bar c)[8]}(x_A,P_\perp) \ ,
\end{eqnarray}
for the color-singlet and octet channels, respectively.
The leading Born cross sections $\sigma_0^{(1,8)}$ are the same
as those calculated in Ref.~\cite{Sun:2012vc}.

\subsection{CEM Calculations}

Similar calculations can be done in the color-evaporation model. The only difference
is that the hard partonic cross sections are calculated in terms of open heavy flavor
pair production with certain invariant mass range,
\begin{eqnarray}
d\sigma=\int dM_{Q\overline{Q}} d\hat{\sigma}(Q\bar{Q}(M_{Q\overline{Q}})+X)f_{M_{Q\overline{Q}}} \; .
\end{eqnarray}
where the invariant mass of the heavy quark pair is restricted in the range of
\begin{equation}
M_\Psi\leq M_{Q\overline{Q}}\leq M_{D\overline{D}} \ ,
\end{equation}
where $M_\Psi$ represents the bound state mass and $M_{D\bar D}$ for the
threshold for the open heavy meson pair. In each case of Charm and Bottom quarks,
we notice that the invariant mass difference between the open heavy meson and
the bound state is very small. We argue that in this model, the heavy quarkonium
production is dominated by the pair production close to the bound state mass range.
In the partonic scattering process, this argument tells us that the momentum fraction
of the heavy quark and antiquark is restricted to $z\approx 1/2$ region. That will
simplify the derivation of the $k_\perp$ factorization formalism.

In the large $N_c$ limit, the heavy quark pair production in $pA$ collisions can be formulated
following the results of Ref.~\cite{Blaizot:2004wv,Dominguez:2010xd}, and we find the leading power contribution
\begin{align}
\frac{d\sigma^{CEM}}{dyd^2P_\perp}\propto &\int dx_{p}\;g_{f}(x_{p})\alpha_S\left[z^2+(1-z)^{2}\right] z(1-z)\frac{S_\perp}{(2\pi)^2}\notag \\
&\times\int\text{d}^2q_1\text{d}^2q_2\delta^{(2)}(P_\perp-q_1-q_2)F_{x_g}(q_1)F_{x_g}(q_2)\frac{(zq_1-(1-z)q_2)^2}{\left[M_{Q}^2+(zq_1-(1-z)q_2)^2\right]^2}.
\end{align}
If we take the limit of small transverse momentum $P_\perp\ll M_Q$,
the above expression can be simplified as,
\begin{equation}
\int\text{d}^2q_1\text{d}^2q_2\delta^{(2)}(P_\perp-q_1-q_2)F_{x_g}(q_1)F_{x_g}(q_2){(zq_1-(1-z)q_2)^2} \ .
\end{equation}
Now, we apply the approximation $z\sim (1-z)\sim 1/2$, and we will find
that the differential cross section of heavy quark pair in the invariant
mass range close to the bound state can be written as
\begin{equation}
\int\text{d}^2q_1\text{d}^2q_2\delta^{(2)}(P_\perp-q_1-q_2)F_{x_g}(q_1)F_{x_g}(q_2){(q_1-q_2)^2}
\propto {\cal F}^{(c\bar c)[8]}\ .
\end{equation}
Clearly, this tells us that the UGD from nucleus is the same as that
in the color-octet contribution in the NRQCD.

The above derivation is consistent with the argument that the CEM and NRQCD
share similar factorization argument for heavy quarkonium production.
Although they differ in the hadronization of the heavy quark pair, but the nuclear
dependence at small-$x$ is the same. This is an important message
for us to use the heavy quarkonium production as a probe to the nuclear
effects at small-$x$.

\section{QCD soft gluon resummation}

An important physics has to be taken care in the low transverse momentum
region of the QCD hard processes. That is the Sudakov double logarithms
in terms of $\ln M^2/P_\perp^2$ from the initial-state gluon shower 
\cite{Berger:2004cc,Sun:2012vc}, 
which has to be resummed to have reliable predictions. 
Recently, there have been studies to show that the Sudakov
resummation can be consistent carried out with the small-$x$ resummation~\cite{Mueller:2012uf}.
In the following, we conjecture that it is the same for heavy quarkonium
production. It will be interesting to have a rigorous calculation to demonstrate
this. We hope to come back to this issue in a later publication.

First, we recall the 
CSS resummation formalism \cite{Collins:1984kg}.  
The resummation part of total cross section can be written as
\begin{equation}
\label{WY} {\frac{d\sigma }{ d^{2}P_{\perp}dy\, \, }}|_{P_\perp\ll M}=
{\frac{1}{(2\pi )^{2}}} \int d^{2}b\, e^{i{\vec{P}_{\perp}}\cdot
{\vec{b}}}{W(b,M,x_{1},x_{2})}\ ,
\end{equation}
where
\begin{eqnarray}
W(b,M^2)=e^{-{\cal S}_{Sud}(M^2,b)}
W(b) \ ,
\label{wpiece}
\end{eqnarray}
with the Sudakov form factor is
\begin{equation}
{\cal S}_{Sud}=\int_{C_0^2/b^2}^{M^2}\frac{d
\mu^2}{\mu^2}\left[\ln\left(\frac{M^2}{\mu^2}\right)
A+B \right]\ .
\end{equation}
$W(b,M,x_{1},x_{2})$ can be written as
\begin{eqnarray}
W(b)&=&\sigma_{0}x f(x,c_0/b) x'f(x',c_0/b)\left(1+\frac{\alpha_s}{\pi}\right) \ ,\label{cgluon}
\end{eqnarray}
where $x=Me^y/\sqrt{S}$, $x'=Me^{-y}/\sqrt{S}$, and the scale $\mu=C_3/b$.

In Ref.~\cite{Mueller:2012uf}, the Sudakov resummation has been extended to the small-$x$
saturation formalism, taking the example of massive color-neutral scalar
particle production in $pA$ collisions. We assume that the general arguments for separating
Sudakov large logarithms and small-$x$ resummation hold for any hard processes,
in particular in the leading double logarithmic approximation~\cite{Mueller:2012uf}. Therefore,
in this paper, we follow the similar formalism, and conjecture that the
Sudakov resummation for heavy quarkonium production can be included in the small-$x$
factorization formalism. We can write
\begin{eqnarray}
\frac{d\sigma^{\rm ({\rm resum})}}{dyd^2P_\perp}|_{P_\perp\ll M}&=&\sigma_0
\int \frac{d^2x_\perp d^2x_\perp'}{(2\pi)^2}e^{iP_\perp\cdot r_\perp}e^{-{\cal S}_{sud}(M^2,r_\perp^2)}
{\cal F}^{(c\bar c)[8]}_{Y=\ln 1/x_a}(x_\perp,x_\perp')\nonumber\\
&&\times xg_p(x,\mu^2={c_0^2}/{r_\perp^2}) ,
\label{resum}
\end{eqnarray}
for the color-octet contribution in NRQCD and CEM model calculations
in the large $N_c$ limit. 
Here, $\sigma_0$ represents the leading order normalization,
$r_\perp=x_\perp-x_\perp'$, ${\cal F}^{(c\bar c)[8]}$ 
denotes the WW-gluon distribution from the nucleus defined in Eq.~(\ref{eq:WWgluon}), 
$g_p(x)$ the integrated gluon distribution from the nucleon. 
The Sudakov form factor ${\cal S}_{sud}$ contains all order resummation, 
and can be calculated perturbatively \cite{Mueller:2012uf}.
For the color-singlet model, we have to replace
${\cal F}^{(c\bar c)[8]}$ by the WW gluon distribution $xG_A^{(1)}(x_\perp,x_\perp')$.

The above result is valid in the limit of $P_\perp\ll M$,
and we have applied the small-$x$ factorization where higher order
in $1/\ln(1/x_a)$ has been neglected as well.
Therefore, the only nuclear effects can be formulated in terms
of the initial state input of the UGDs.  
When $P_\perp \sim Q_s(A) \sim M$ and the condition in Eq.~(\ref{eq:fac-condition}) is
satisfied, the nuclear multiple scattering based effective $k_\perp$-factorization 
for heavy quarkonium production in $pA$ collisions in the forward region could be still valid, 
and a matching $Y$-term could be added to the resummed contribution  
\cite{Collins:1984kg} 
to extend the validity of Eq.~(\ref{resum}) to the region where $M\sim Q_s(A)$. 

\section{Nuclear Suppression}

When applying Eq.~(\ref{resum}) to calculate the nuclear suppressions
for heavy quarkonium production in $pA$ collisions, we need input of the
UGDs. 
First, we find that the UGD involved in heavy quarkonium production
can be calculated in the MV model,
\begin{eqnarray}
xG_A^{(1)}(r_\perp)&=&\frac{S_{\perp }}{\pi ^{2}\alpha _{s}}\frac{
N_{c}^{2}-1}{N_{c}}\frac{1}{r_{\perp }^{2}}\left( 1-e^{-\frac{r_{\perp
}^{2}Q_{s}^{2}(A)}{4}}\right) \ ,\\
{\cal F}^{(c\bar c)[8]}(r_\perp)&=&\frac{S_{\perp }}{\pi ^{2}\alpha _{s}}\frac{
N_{c}^{2}-1}{N_{c}}\frac{Q_{s}^2(A)}{4} e^{-\frac{r_{\perp
}^{2}Q_{s}^{2}(A)}{4}} \ ,\label{fcc}
\end{eqnarray}
where $Q_{s}(A)$ is the saturation scale in the adjoint representation.
At small dipole size, the above UGDs are the same, and both
proportional to the integrated gluon distributions,
\begin{equation}
Q_{s}^2(A)
=\frac{2N_c^2}{N_c^2-1}A^{1/3}Q_{sp}^2\approx
\frac{2N_c^2}{N_c^2-1}\frac{2\pi^2\alpha_s}{N_cS_\perp}xG_A(x,1/r_\perp),
\end{equation}
where the last quantity is the integrated gluon distribution.  It is easy to find that this gives the
correct normalization after integrating over the transverse momentum $P_\perp$. From the calculations in Sec.~II, we know that the
heavy quarkonium production at low $P_\perp$ is dominated ${\cal F}^{(c\bar c)[8]}$
gluon distribution (coming from  
the color-octet channel in NRQCD factorization or quark pair in
large $N_c$ in the the CEM model). Therefore, our following discussions
will focus on this part of contribution

With above initial parameterization for the UGDs, we will be
able to calculate the transverse momentum distribution of heavy quarkonium
in $pA$ collisions, and study the nuclear suppression factor. 
Although the precise predictions of the nuclear suppression
depend on the numeric calculations for heavy quarkonium production
in both $pp$ and $pA$ collisions, a number of interesting features can be
derived from the above formulas. First, the suppressions when $P_\perp\to 0$ 
is less sensitive to the detailed shape of the UGDs.  
This is because, without $e^{iP_\perp\cdot r_\perp}$ when $P_\perp=0$, 
the differential cross section depends on the integral over full range of $r_\perp$, 
rather than the limited region, $r_\perp \lesssim 1/P_\perp$, 
as shown in Eq.~(\ref{resum}).  
Second, since higher saturation scale broadens the $P_\perp$-distribution 
(as can be seen from Eq.~(\ref{fcc})), we will have less suppression
at higher $P_\perp$. 
This phenomena is generic \cite{Guo:1999wy} and 
has been seen from experiments. 
That is, the $P_\perp$-distribution of heavy quarkonium production 
in the forward region of $pA$ collisions is very sensitive to the shape of the UGDs.

Therefore, heavy quarkonium production at low transverse momentum 
in the forward region of $pA$ collisions has a strong advantage to probe
the saturation physics at small-$x$ in a large nucleus.  We find that 
the nuclear suppression in this region is not very sensitive to the details 
on how does a produced heavy quark pair transform into an physical quarkonium 
since the NRQCD and CEM yield the same suppression factor in low $P_\perp$ region.

\section{Summary}

In this paper, we have investigated the heavy quarkonium production
in the forward region of $pA$ collisions in the low transverse momentum region 
in the small-$x$ factorization formalism. Following previous example, we
developed an effective $k_\perp$-factorization, where the new combination
of the un-integrated gluon distributions dominate the production of heavy
quarkonium in $pA$ collisions. More importantly, predictions calculated using
both NRQCD and CEM depend on the same UGD in the large $N_c$ limit, which
will yield a universal nuclear suppression factor for heavy quarkonium production.

In Ref.~\cite{Sun:2012vc}, heavy quarkonium production at low $P_\perp$
has been applied to a wide energy range of the $pp$ collisions,
from fixed target to colliders such as the LHC and Tevatron. It will be
interesting to extend the numerical calculations to the $pA$ collisions, 
by applying the results derived in this paper. 
We will leave this to a future publication.
We notice that the small-$x$ calculation of heavy quarkonium
production in the CEM has recently been published~\cite{Fujii:2013gxa}. 

We thank Al Muller for stimulating discussions.
This work was supported in part by the U.S. Department of Energy under the
contracts DE-AC02-98CH10886 and DE-AC02-05CH11231.

{\it Note added}: when this paper was finished, we noticed a preprint~\cite{Kang:2013hta} about
the same topic. These two calculations agree with each other in the kinematics
we are interested in this paper.

\appendix
\section{Basics in NRQCD Calculations}

The spin projectors for outgoing heavy quarks momenta $Q = P/2+q$
and ${\overline Q }= P/2-q$, are given by

\begin{eqnarray}&&{\mit \Pi}_{0}
     = {1\over{\sqrt{8m^3}}} \left({{\slash{P}}\over 2} - \slash{q} - m\right)
       \gamma_5 \left({{\slash{P}}\over 2} + \slash{q} + m\right)\, ,
\label{proj_00}
\\
&&{\mit \Pi}_{1}^{\rho}
     = {1\over{\sqrt{8m^3}}} \left({{\slash{P}}\over 2} - \slash{q} - m\right)
       \gamma^\rho \left({{\slash{P}}\over 2} + \slash{q} + m\right)\, .
\label{proj_1Sz}\end{eqnarray} 
The colour singlet or octet state
content of a given state can be projected out by contracting the
amplitudes with the following operators
\begin{eqnarray}&&{\cal
C}_1 = {{\delta_{ij}}\over{\sqrt{N_c}}}\qquad{\rm for~the~singlet}
\label{proj_sing}
\\
&&{\cal C}_8 = \sqrt{2} T_{ij}^c\qquad{\rm for~the~octet} .
\label{proj_oct}\end{eqnarray}

The projection on a state with orbital angular momentum $L$ is
obtained by differentiating $L$ times the spin-colour projected
amplitude with respect to the momentum $q$ of the heavy quark in the
$QQ$ rest frame, and then setting $q$ to zero. We shall only deal
with either $L=0$ or $L=1$ states, for which the amplitudes take the
form
\begin{eqnarray}&&{\cal A}_{S=0,L=0} = Tr\left.\left[{\cal C}\,{\mit  \Pi}_0\,
{\cal A}\right]
\right|_{q=0}\qquad\qquad\qquad\,{\rm Spin}\;{\rm  singlet }~S~{\rm states}\\
&&{\cal A}_{S=1,L=0} = \epsilon_\rho
          Tr\left.\left[{\cal C}\,{\mit  \Pi}_1^\rho\,
     {\cal A}\right]\right|_{q=0}\qquad\qquad\;\;\,\,\,{\rm Spin}
     \;{\rm triplet }~S~{\rm states}\\
&&{\cal A}_{S=0,L=1} = \epsilon_\nu {d\over{q_\nu}}
Tr\left.\left[{\cal C}\, {\mit \Pi}_0 {\cal
A}\right]\right|_{q=0}\quad\qquad\;\;\,{\rm Spin}\;
{\rm singlet }~P~{\rm states}\\
&&{\cal A}_{S=1,L=1} = {\cal E}_{\rho\nu} {d\over{ q_\nu}}
Tr\left.\left[{\cal C}\, {\mit \Pi}_1^\rho {\cal
A}\right]\right|_{q=0}\quad\qquad{\rm Spin}\; {\rm triplet }~P~{\rm
states}. \end{eqnarray}
\section{Color-octet ${}^3P_J^{(1,8)}$ Channel}
For $P$-wave states, which are different from $S$-wave states, we have to do the expansion of amplitude
to order $q^\nu$ as shown in Eqs.(A7, A8). After a lengthy evaluation, we find that Fig.~2(a)
contribution can be written as
\begin{eqnarray}
\mathcal {M}^{\mu\rho\nu}( {}^3P_J^8)_{\rm fig.2(a)}
=\Xi^{\rho\nu\mu}[k_{g1\perp},k_{g2\perp}]\frac{1}{\sqrt{4m_c^3}}2\pi\delta(k_g^-)(\textrm{Tr}[T^bU(x_\perp)T^aU^\dag(x_\perp)]),
\end{eqnarray}
where
\begin{eqnarray}
\Xi^{\rho\nu\mu}[k_{g1\perp},k_{g2\perp}]\=
\frac{d}{dq^\nu}\textrm{Tr}[\slash{n}\frac{\mathbbm{i}}{\slash{p}+\slash{q}-\slash{k}_{g1}-m_c}\gamma^\mu
\frac{\mathbbm{i}}{-\slash{p}+\slash{q}+\slash{k}_{g2}-m_c}\slash{n}
(\slash{p}-\slash{q}-m_c)\gamma^\rho(\slash{p}+\slash{q}+m_c)]\nn .
\end{eqnarray}
Similar expression is found for the color-singlet channel,
\begin{eqnarray}
\mathcal {M}^{\mu\rho\nu}( {}^3P_J^1)_{\rm fig.2(a)}
=\Xi^{\rho\nu\mu}[k_{g1\perp},k_{g2\perp}]\frac{1}{\sqrt{4m_c^3}}2\pi\delta(k_g^-)(\textrm{Tr}[U(x_\perp)T^aU^\dag(x_\perp)])
\end{eqnarray}
\begin{eqnarray}{\cal M}_{S=1,L=1} = {\cal E}_{\rho\nu} {d\over{ q_\nu}}
Tr\left.\left[{\cal C}\, {\mit \Pi}_1^\rho {\cal
M}\right]\right|_{q=0}\quad\qquad{\rm Spin}\; {\rm triplet }~P~{\rm
states} .
\end{eqnarray}

 We define
\begin{eqnarray}
\Pi^{\rho\nu}=-g^{\rho\nu}+\frac{p^\rho p^\nu}{m_c^2} ,
\end{eqnarray}
and the sum of polarization for a ${}^3P_J$ state, which can be cast into the following
forms
\begin{eqnarray}
\varepsilon^{(0)}_{\rho\nu}\varepsilon^{(0)}_{\rho'\nu'}&=&\frac{1}{D-1}\Pi^{\rho\nu}\Pi^{\rho'\nu'}\\
\sum\varepsilon^{(1)}_{\rho\nu}\varepsilon^{(1)}_{\rho'\nu'}&=&\frac{1}{2}\left[\Pi^{\rho\rho'}\Pi^{\nu\nu'}-\Pi^{\rho\nu'}\Pi^{\rho'\nu}\right]\\
\sum\varepsilon^{(2)}_{\rho\nu}\varepsilon^{(2)}_{\rho'\nu'}&=&\frac{1}{2}\left[\Pi^{\rho\rho'}\Pi^{\nu\nu'}+
\Pi^{\rho\nu'}\Pi^{\rho'\nu}\right]-\frac{1}{D-1}\Pi^{\rho\nu}\Pi^{\rho'\nu'},
\end{eqnarray}
and
\begin{eqnarray}
\varepsilon^{(0)}_{\rho\nu}\mathcal
{M}({}^3P_0^8)^{\mu\rho\nu}=\frac{16 \mathbbm{i} \left(k_{g2}^{\mu
}- k_{g1}^{\mu }\right) \left(-12 m_c^2+(k_{g2}-k_{g1})^2\right)
n\cdot p}{\left(-4
m_c^2-(k_{g2}-k_{g1})^2\right)^2}\frac{1}{\sqrt{4m_c^3}}2\pi\delta(k_g^-)(\textrm{Tr}[T^bU(x_\perp)T^aU^\dag(x_\perp)])\nn .\\
\end{eqnarray}
At the leading order of $k_\perp$, we find 
\begin{eqnarray}
\varepsilon^{(0)}_{\rho\nu}\mathcal
{M}({}^3P_0^8)^{\mu\rho\nu}=\frac{ \mathbbm{i} \left(k_{g2}^{\mu }-
k_{g1}^{\mu }\right) \left(-12 \right) n\cdot p}{
m_c^2}\frac{1}{\sqrt{4m_c^3}}2\pi\delta(k_g^-)(\textrm{Tr}[T^bU(x_\perp)T^aU^\dag(x_\perp)]).
\end{eqnarray}
Similarly, we can obtain
\begin{eqnarray}
\varepsilon^{(0)}_{\rho\nu}\mathcal
{M}({}^3P_0^1)^{\mu\rho\nu}=\frac{16 \mathbbm{i} \left(k_{g2}^{\mu
}- k_{g1}^{\mu }\right) \left(-12 m_c^2+(k_{g2}-k_{g1})^2\right)
n\cdot p}{\left(-4
m_c^2-(k_{g2}-k_{g1})^2\right)^2}\frac{1}{\sqrt{4m_c^3}}2\pi\delta(k_g^-)(\textrm{Tr}[U(x_\perp)T^aU^\dag(x_\perp)]), \nn\\
\end{eqnarray}
\begin{eqnarray}
\varepsilon^{(0)}_{\rho\nu}\mathcal
{M}({}^3P_0^1)^{\mu\rho\nu}=\frac{ \mathbbm{i} \left(k_{g2}^{\mu }-
k_{g1}^{\mu }\right) \left(-12 \right) n\cdot p}{
m_c^2}\frac{1}{\sqrt{4m_c^3}}2\pi\delta(k_g^-)(\textrm{Tr}[U(x_\perp)T^aU^\dag(x_\perp)]),
\end{eqnarray}
and
\begin{eqnarray}
&&\mathcal {M}({}^3P_2^8)^{\mu\rho\nu}\mathcal
{M}({}^3P_2^8)^{\mu'\rho'\nu'}\varepsilon^{(2)}_{\rho\nu}\varepsilon^{(2)}_{\rho'\nu'}\nn\\&=&
\left(\frac{32 g^{\mu
   \mu'} (k_{g1}-k_{g2})\cdot(k'_{g1}-k'_{g2}) (n\cdot p)^{2}}
   {m_c^4}\right)\left(\frac{1}{\sqrt{4m_c^3}}2\pi\delta(k_g^-)(\textrm{Tr}[T^bU_{k_{g1}}(x_{1\perp})T^aU_{k_{g2}}^\dag(x_{2\perp})])\right)\nn\\
   &\times&\left(\frac{1}{\sqrt{4m_c^3}}2\pi\delta(k'{}_g^-)(\textrm{Tr}[T^bU_{k'_{g1}}(x'_{1\perp})T^aU_{k'_{g2}}^\dag(x'_{2\perp})])\right)^\dag .
\end{eqnarray}

\end{document}